# The Effect of Data Types' on the Performance of Machine Learning Algorithms for Financial Prediction

Hulusi Mehmet Tanrikulu, Hakan Pabuccu

**Abstract**

Forecasting cryptocurrencies as afinancial issue is crucial as it provides investors with possiblefinancial benefits. A small improvement in forecasting performance can lead to increased profitability; therefore, obtaining a realistic forecast is very important for investors. Successful forecasting provides traders with effective buy-or-hold strategies, allowing them to make more profits. The most important thing in this process is to produce accurate forecasts suitable for real-life applications. Bitcoin, frequently mentioned recently due to its volatility and chaotic behavior, has begun to pay great attention and has become an investment tool, especially during and after the COVID-19 pandemic. This study provided a comprehensive methodology, including constructing continuous and trend data using one and seven years periods of data as inputs and applying machine learning (ML) algorithms to forecast Bitcoin price movement. A binarization procedure was applied using continuous data to construct the trend data representing each input feature trend. Following the related literature, the input features are determined as technical indicators, google trends, and the number of tweets. Random forest (RF), K-Nearest neighbor (KNN), Extreme Gradient Boosting (XGBoost-XGB), Support vector machine (SVM) Naive Bayes (NB), Artificial Neural Networks (ANN), and Long-Short-Term Memory (LSTM) networks were applied on the selected features for prediction purposes. This work investigates two main research questions: i. How does the sample size affect the prediction performance of ML algorithms? ii. How does the data type affect the prediction performance of ML algorithms? Accuracy and area under the ROC curve (AUC) values were used to compare the model performance. A t-test was performed to test the statistical significance of the prediction results. When examining the continuous data models, the ANN model showed the best performance with the highest accuracy and outperformed the otherfive ML models for complete and one-year data sets. Also, the SVM model showed significantly similar performance to the best ANN model. The ANN and SVM models showed the best performance for a one-year data set with lower accuracy values when compared to the complete data set used models. Similarly, the LSTM model performed best with a 0.957 accuracy value and outperformed The ANN, KNN, and NB models for trend data. Also, the SVM, NB, and XGB models showed significantly equal performance on trend data. The detailed experimental analysis indicates that the proposed methodology can forecast return movement efficiently and accurately. Trend data constructed for this study improved the performance of ML models.

# 1 Introduction

With the introduction of the internet into our lives in the past thirty years, our daily habits have undergone a rapid digital transformation, and new virtual realities beyond physical spaces and borders have become a part of our lives. As part of this transformation, business models that interact with each other throughout the day have started to develop worldwide, independent of the concept of working hours. Parallel to the development of technology and the internet, the concept of money has also been digitized to a large extent and has become numbers connected to bank accounts. In addition to all these developments, in 2009, a completely digital currency called Bitcoin (BTC), which is independent of banking systems, not connected to any state or center, not physically produced, and can be obtained in a limited number by arranging digital blocks in a certain order has entered our lives as a new medium of exchange Delfabbro et al. (2021).

Cryptocurrency markets have become a rapidly expanding global market after the discovery of the BTC. Cryptocurrencies are used for cross-border payments and financial investments, making them a valuable digital and safe haven asset Kim et al. (2021). Parallel to the development of the digital world, data analysis tools, and methods are also developed in the same way. Researchers and investment experts try to develop better estimation tools by combining fundamental and technical analysis methods and using traditional statistical and machine learning techniques for stock market prediction. The fact that Bitcoin has been frequently traded as an investment tool, especially with the pandemic effect, has increased the work done in this field Kim et al. (2020).

Predicting the price of Bitcoin is a complex and challenging task due to the various technical and statistical challenges inherent in financial time series prediction. One of the biggest challenges is the risk of model overfitting, where a model is fitted too closely to training data and cannot generalize well to new, unseen data. This problem is exacerbated by the high volatility and non-linearity of Bitcoin price data, which makes it vital to use the right regularisation techniques and model evaluation methods to avoid overfitting. Khosravi and Ghazani (2023)

There are over 22,000 different cryptocurrencies with a market capitalization of over $1.95 Trillion as of February 2024 CoinMarketCap (2024). Among all these cryptocurrencies, Bitcoin, founded in 2009, has been the most popular

since its inception. It conquered the cryptocurrency market with a 42% share and a market capitalization of over $1 trillion.

This article aims to make the following contributions:

The problem is formalized as a classification to predict the BTC price movements or trends. The proposed models investigated the effect of continuous and trend data sets on prediction performance and the effect of the amount of data samples. It also discussed the importance of Google trends and tweet data on BTC price movements.

The remainder of this paper is organized as follows: Section 2 presents the "Related works". Section 3 explains the data sets and applied methodology. Section 4 presents the experimental results and the performance comparison of different prediction models. Finally, Section 5 finishes the discussions and conclusions.

## 2  Related works

Financial time series prediction techniques are organized into four categories, namely statistical methods, machine learning, pattern recognition, and sentiment analysis by Shah et al. (2019). The financial time series prediction problems can also be categorized into classification and regression. In this section, statistical methods and machine learning models are taken into consideration within the scope of this research. Statistical models make some simplifying assumptions to obtain some theoretical guarantees and thus have a simple structure and potentially lover performance compared to machine learning techniques. Machine learning models are more general, make fewer simplifying assumptions, and offer better model-fitting capabilities.

One of the first studies on Bitcoin price movement prediction was made by Shah and Zhang (2014) using the Bayes Linear Regression model. Bitcoin price data from February to July 2014 were taken, and the k-mean clustering ML method was applied to over 200 million raw data points. Madan et al. (2015) applied machine learning algorithms to predict BTC price movements using two different datasets consisting of BTC daily closing values with 10-second and 10-minute frequency. The authors applied the Generalized Linear Model (GLM), SVM, and RF models. Results were evaluated using Sensitivity, Specificity, Precision, and Accuracy metrics. As a result of the study, the GLM and RF models provided better results than the SVM.Huang et al. (2019)used 124 financial technical indicators from the daily data of Bitcoin between 2012 and 2018 with 2168 samples. Classification and Regression Trees (CART) were used as a machine learning method to forecast bitcoin prices. In this study, the proposed model has strong (out-of-sample) predictive power for narrow ranges of Bitcoin's daily returns and evidence that technical analysis is useful in a market like Bitcoin, whose value is primarily driven by non-fundamental factors. Mangla et al. (2019) used ARIMA, SVM, LR, and RNN methods to forecast Bitcoin price movements. The dataset consists of Bitcoin one-hour intervals between 2015-2019. In this study, the performances of ML algorithms were tested, and attempts were made to increase their performance. As a result of the study, among the four methods, ARIMA performs well for next-day predictions but performs poorly for longer terms, like given the last few days' price predictions for the next 5-7 days' prices. RNN performs consistently up to 6 days. Chen et al. (2020) used two separate data sets. The first dataset consists of Bitcoin daily, and the other dataset consists of 5 minutes of Bitcoin price movements between 2017-2019. Block size, Hash rate, Excavation Rate, number of transfers, indicators of blockchain and mining information, Google trend searches, transaction volume, and gold prices were used as inputs. LR and Linear Discriminant Analysis (LDA), RF, XGB, Quadratic Discriminant Analysis, SVM, and LSTM were applied. Accuracy, Precision, Recall, and F1-Score were calculated to compare the performance of the selected algorithms. On the daily close dataset, LR gave the best results with 66% Accuracy, while XGB remained at 48.3%. For the 2nd data set, LSTM gave the best results with 67.2%, while LR was 54.5% and LDA 51.5% level of accuracy. Lahmiri and Bekiros (2020) employ three different sets of models: statistical machine learning SVR and GRP, algorithmic models RT and KNN, and artificial neural network topologies FFNN, BRNN and RBFNN. 5-minute intervals from the beginning of 2016 to March 2018 were taken as datasets. For each machine learning model, the first 80% of the dataset is employed for training while the remaining 20% is for testing. The optimal parameter values for SVR, GRP, and kNN are found via Bayesian optimization. Based on diverse performance metrics, BRNN renders an outstanding accuracy in prediction, while its convergence is unhindered and remarkably fast. Pabuçcu et al. (2020) used daily Bitcoin data between 2008-2019. The number of samples is 1935. The authors used an additional 100 days after June 2020 as a validation set. SVM, ANN, NB, and RF techniques were used, and LR was used for benchmarking. Accuracy, MAE, and RMSE were calculated to evaluate the model performance and conduct a t-test for statistical significance. While RF showed the highest prediction performance for the continuous dataset, the NB was the lowest. The ANN and NB were the best and the worst prediction models for trend data sets, respectively. Furthermore, the trend datasets improved the overall prediction performance in all estimated models. Chen et al. (2021) used economic (gold, oil, SP500, exchange rates, etc.) and technological (Hasrate, average transaction fees, Tweet, Google trends, etc.) data as input in their studies. Four different data sets were created by determining the important price rise and fall ranges of Bitcoin between 2011 and 2018, and the last 2 months of data were selected as the prediction set. ANFIS, SVR, ARIMA, and LSTM machine learning algorithms were used as methods. In the study, the LSTM model gave the best MAE, MAPE, and RMSE test results. The results demonstrated that using the economic and technology determinants effectively predicted the Bitcoin exchange rate. Sossi-Rojas et al. (2023) tried to predict Bitcoin price movements using LSTM, BiLSTM, GRU, BiGRU, and LightGBM techniques. They created a data set consisting of Hash Rate, Exchange Rate, Google Trend, Fear Greed, and Bitcoin prices as features between 2013-2022. RMSE, MSE, MAE, and directional accuracy(DA) were used to evaluate the model performance. BiLSTM showed the best performance in terms of DA.

Goutte et al. (2023) In their study, they obtained hourly Bitcoin data between 2017 and 2022 using Binance API. They used SMA, RSI, MACD, WR, STOC, and MFI as technical indicators. Seven machine learning algorithms were used: XGBoost, LightGBM, MLP, GRU, LSTM, and CNN. Unlike other studies, candlestick charts were converted into a discrete data structure and interpreted using it. Analyzes were made using GRU, CNN, LSTM, MLP, Logistic, LGBM, and XGBoost methods, and GRU showed the best performance. Rahaman et al. (2022) tried to develop a new indicator model to interpret bitcoin prices. They obtained data from various sources using the BeautifulSoup library, as we used in our study. They aimed to produce a buy-sell signal by taking daily tweets with the Phython tweepy plugin, conducting sentiment analysis with TextBlob, and interpreting them with the FBProphet module. Their study's most important finding was they could profit by USD 1,000.71 in just 10 days with RMSE: 1480.58.

## 3 Data preparation

The scope of the research was limited to the Bitcoin price movement prediction problem. Other crypto assets are not included in this study. In this research, three different types of information were used from the original data. i. Historical prices from Yahoo Finance contain open, close, highest (High), lowest (Low), and transaction volume (Volume) daily values of Bitcoin on a USD basis using Python Yahoo Finance API. ii. Technical indicators: Simple Moving Averages (SMA), Weighted Moving Average (WMA), Exponential Moving Average (EMA) Average True Range (ATR), Bollinger Bands (Boll), Momentum (Mom), Stochastic indicators (K and D), Relative Strength Index (RSI), Moving Average Convergence and Divergence (MACD), Larry Williams(LW), Accumulation Distribution Indicator (A/D) were calculated using Python Technical Analysis Library (TA-Lib). Following the finance literature, it is decided to use these technical indicators, which are very common and powerful, as input features for selected prediction ML techniques. iii. The daily tweets about Bitcoin and Google search trend data were obtained from https://bitinfocharts.com/ with the Python BeautifulSoup library.

By using these three data sources, continuous and trend data sets were constructed to provide feature sets to the selected ML techniques as inputs. It was provided detailed information about the mentioned data sets constructed in the next section.

The complete dataset contains data between 17/09/2014-10/06/2021, and the reduced dataset contains data between 10/06/2020-10/06/2021. The validation set contains the data between 11/06/2021-09/08/2021 to validate the estimated models for 30-60 days, respectively.

The output is defined as trend data, representing the up-down movements of the daily closing price. If any sample point is 0, it means a downward trend, and if it is 1, it means an upward trend. The output data was also used by dividing the training and testing sets at the same rate. For the classification problem, a binary output was calculated using (Eq. 1.). The up and down movement was coded as $\{0, 1\}$ for the prediction models. This is a very common process to represent the trend of the prices and returns, or in other words, economic profits or losses.

$$y_t = \begin{cases} 1, & \text{if } return_t > return_{t-1} \\ -1, & \text{else} \end{cases} \quad (1)$$

## 3.1 Continuous data preparation

The complete dataset, containing 17 columns and 2455 rows using technical indicators and google-tweet, was normalized to the 0-1 range, and a continuous data set was created. The reduced data set was created by trimming the last 365 days of the complete data set. The definitions and calculation procedures of the technical indicators used in the datasets are as follows:

**Accumulation Dispersion Indicator (A/D) (ADOSC):** It is a momentum indicator that examines the relationship between price change and volume. It is interpreted as the sustainability of the price movement.

**Bollinger Bands:** Bollinger Bands, one of the most frequently used technical analysis indicators, are bands that show volatility by being placed above and below the moving average. Volatility is a metric that depends on standard deviation and measures the fluctuation level of selected features. There are three pieces of data in Bollinger Bands. Generally, the middle band is obtained from the 20-day moving average, and the upper and lower bands are obtained by moving the 20-day moving average up and down by two standard deviations.

**Exponential Moving Average (EMA):** It is an indicator where the weighting process is made according to recent price movements by taking the average of the n-time price. Weighting is used to reduce the delay rate.

**Moving Average Convergence/Divergence (MACD):** It is a momentum indicator and shows the relationship between two different price moving averages. They are generally obtained by subtracting the 26-day exponential moving average from the 12-day exponential moving average and are used for trend following.

**Simple Moving Average (SMA):** SMA is calculated by taking the n-day closing average of the n-time price. It is used to resolve fluctuations in data.

**Weighted Moving Average (WMA):** It is an indicator calculated by averaging the n-time price according to the determined weights.

**Relative Strength Index (RSI):** RSI is calculated by comparing the closing price in the selected time unit with the previous closing values. It is an indicator used to predict the direction of short-term and medium-term trends. It is generally taken in 14-day periods and produces a buy signal when it falls below 30 and a sell signal when it rises above 70.

If $A_n - A_n - 1$ is positive, it is defined as Positive Closing (PC), and if it is negative, it is defined as Negative Closing (NC). This process is applied for the entire specified period.

**Average True Range (ATR):** ATR is calculated by averaging the highest and lowest difference of each movement in price. It is used to measure the volatility of the market.

**Larry Williams (LW):** LW is used to detect overbought and oversold areas. If the indicator slope is above -20, it means overbought, and if it is at -80, it means oversold. It is generally calculated for 14 days.

**Momentum (Mom):** It is a technical indicator that measures and shows the strength or speed of a price movement. The MOM indicator compares the latest price with a predetermined price and measures the speed of price change. Traders choose whether price momentum will increase/decrease to determine entry and exit points.

**Stochastic %K %D Oscillator (Stoch):** indicator shows the closeness of high and low values to the current price within a certain period.

Table 1: Indicator Formulas

| Indicator | Formula |
|---|---|
| A/D | $A/D = \frac{H_t - C_{t-1}}{H_t - L_t}$ |
| Bollinger Bandsup | $UPPERBB = MA + D\sqrt{\frac{\sum_{i=1}^{n}(y_j - MA)^2}{n}}$ |
| Bollinger Bands Low | $LOWERBB = MA - D\sqrt{\frac{\sum_{i=1}^{n}(y_j - MA)^2}{n}}$ |
| EMA | $EMA = A_t * k + EMA_{t-1} * (1 - k)$ |
| MACD | $MACD = EMA_{12} - EMA_{26}$ |
| SMA | $SMA = \frac{(A_1 + A_2 + ... + A_n)}{n}$ |
| WMA | $WMA = \frac{\sum_{t=1}^{n} W_t * V_t}{\sum_{t=1}^{n} W_t}$ |
| RSI | $RS = \frac{Average(PC)}{Average(NC)} \qquad RSI = 100 - \frac{100}{1+RS}$ |
| ATR | $ATR = \frac{1}{n} \sum_{i=1}^{n} TR_i$ |
| LW | $LW = \frac{H_n - C_t}{H_n - L_n} * 100$ |
| MOM | $Mom = \frac{C_t}{C_{t-n}} * 100$ |
| StoK - StoD | $StoK = \frac{C_t - L_n}{H_n - L_n} * 100 \qquad StoD = \frac{H_n}{L_n} * 100$ |
| $A$ = Closing price | $t$ = Day of calculation |
| $k = 2 \div$ (total time period $+ 1$) | $A_n$ = closing price at time n |
| n = total time period | $V_t$ = Actual Value at time t |
| $W_t$ = Weighting Value at time t | $A_n$ = closing price at time n |
| $H_t$ = highest price at time t | $L_t$ = Lowest price at time t |
| $TR_i$ = Specified actual range | $H_n$ = highest price in time interval n |
| $C_t$ = closing price at time t | $L_n$ = Lowest price in time interval n |

In addition to technical indicators, it is used tweets and Google trend data about Bitcoin. The purpose of using Google Trends and Twitter tweets in the study is that both factors reflect public attention. However, there are differences between these data. While Google Trends shows the public's intention to learn about Bitcoin, tweets represent the public's intention to discuss Bitcoin. In our study, we observed the effect of the change in these data on the change in the direction of price movement.

Descriptive statistics are presented as follows in Table 2.

Table 2: Descriptive statistics.

| Method | Minimum | Maximum | Mean | St. Dev. |
|---|---|---|---|---|
| SMA | 222.873 | 60050.151 | 7706.481 | 11654.864 |
| WMA | 219.543 | 60904.757 | 7738.658 | 11692.165 |
| EMA | 225.270 | 60397.746 | 7704.597 | 11626.316 |
| ATR | 3.556 | 5003.094 | 450.985 | 833.745 |
| Momentum | -21229.875 | 14372.156 | 148.338 | 2094.090 |
| Stochastic %K | 0 | 100 | 56.147 | 29.923 |
| Stochastic %D | 2.714 | 99.199 | 56.148 | 28.172 |
| Boll-Upper | 211.193 | 66836.082 | 8288.583 | 12601.241 |
| Boll-Middle | 200.878 | 62382.740 | 7776.278 | 11759.605 |
| Boll-Lower | 161.955 | 60159.319 | 7263.974 | 10941.897 |
| RSI | 9.920 | 94.302 | 54.293 | 14.288 |
| LW | -100 | 0 | -43.853 | 29.923 |
| A/D | -53660.135 | 119838.185 | 5269.312 | 14876.947 |
| Tweet | 7300 | 363566 | 37187.716 | 28701.176 |
| Google Trend | 5.268 | 616.867 | 48.825 | 60.356 |

## 3.2 Trend data preparation

A discretization procedure is applied to construct a trend data set using the original continuous data set. The mentioned procedure was presented in table 3. The trend data set was formed from the data coded as 1 for the uptrend and 0 for the downtrend. The last 365 rows of the trend and continuous data sets constitute the reduced sets to test the effect of sample size on prediction performance.

Table 3: Discretization procedure for trend data

| Data | Value 1 (Upward Trend) | Value 0 (Downward Trend) |
|---|---|---|
| SMA | If <Closing price | If >Closing price |
| WMA | If <Closing price | If >Closing price |
| EMA | If <Closing price | If >Closing price |
| ATR | If >previous day | If <previous day |
| Bollinger Bands | If Closing price closer to lower band | If Closing price closer to upper band |
| Momentum | If positive | If negative |
| Stochastic %K | If >previous day | If <previous day |
| Stochastic %D | If >previous day | If <previous day |
| Larry Williams | If >previous day | If <previous day |
| RSI | If <30 or 30-70 and >previous day | If >70 or 30-70 and <previous day |
| MACD | If >previous day | If <previous day |
| A/D | If >previous day | If <previous day |
| Tweet | If >previous day | If <previous day |
| Google Trend | If >previous day | If <previous day |

While constructing the trend data it is applied a highly similar procedure for each feature to calculate and code up-down movement. Since the momentum value indicates the sustainability of the trend, it is assigned a value of 1 if it is positive and 0 if it is negative. Since the stochastic indicators %K, %D, and Larry Williams(LW) indicator that shows the rate of price change and overbought; if it is higher than the previous day means trend is upwards and value 1, if it is lower means trend is downwards and value 0 assigned. While interpreting the RSI, if the value is less than 30, the trend is upwards and value 1 assigned. If it is between 30-70 and if it is greater than the previous day value 1 assigned, if not value 0 assigned. If the RSI is greater than 70 means trend goes downwards and value 0 assigned. For MACD, if it is higher than the previous day value 1 ,if it is lower, a value of 0 is assigned. The A/D indicator is assigned in the same way, 1 if the Oscillator value is higher than the previous day's value, and 0 if it is lower. The number of tweets sent per day and Google Search Trend are also compared with the previous day and if higher value 1, and if lower value 0 assigned.

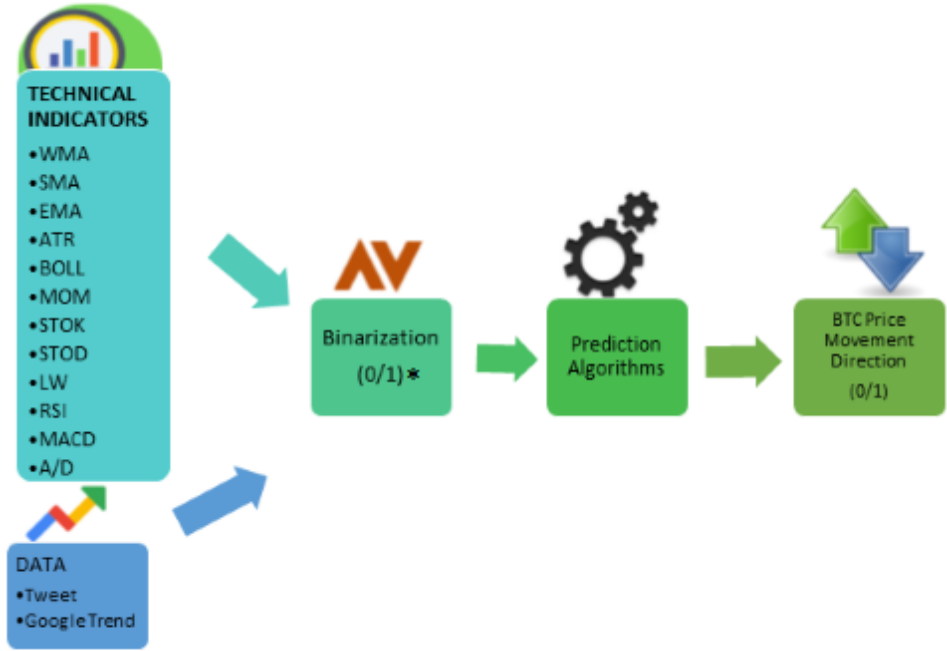

Figure 1: prediction mechanism

During the preparation of the 2455-line data set, the first 13-day values for RSI, EMA, SMA, WMA, StoK, StoD, the first 9-day values for ADOSC, the first 3-day values for LW, and the first 33-day values for MACD could not be calculated due to the formulation. Google trend data has no missing values, while tweet data has 56. Those NaN values are replaced to 0 with fillna(0) command.

### 3.2.1 Verification Method Analyzes

Train-Test Split and Cross-Validation methods were tested to choose the most suitable validation method for our data set. When we tested the accuracy of our trend data set with the K-nearest neighbor method with random state settings and the trend-test split method with different constants between 1-10, we achieved an average accuracy of 0.94. 4. It has been observed that the accuracy rate changes with each constant in the data. This may mean we encounter different results each time the method is tested.

Table 4: Training-Test Decomposition Accuracy Change

| Random State | Accuracy |
|---|---|
| 1 | 0.941368 |
| 2 | 0.939739 |
| 3 | 0.926710 |
| 4 | 0.947883 |
| 5 | 0.941368 |
| 6 | 0.959283 |
| 7 | 0.956026 |
| 8 | 0.928339 |
| 9 | 0.941368 |
| 10 | 0.923453 |
| **Mean** | **0.940554** |

According to the data in the classification report5, when the Train-Test Split method use with 25% split, 614 of the 2455 data set could be used for testing, and the average accuracy was 0.94.

Table 5: Train-Test Split Classification Report

|  | Precision | Recall | F Score | Support |
|---|---|---|---|---|
| 0 | 0.94 | 0.94 | 0.94 | 282 |
| 1 | 0.95 | 0.95 | 0.95 | 332 |
| Accuracy |  |  | 0.94 | 614 |
| Makro Acc. | 0.94 | 0.94 | 0.94 | 614 |
| Weighted avg. | 0.94 | 0.94 | 0.94 | 614 |

In the cross-validation method, our data set is divided by the number of layers we have determined, each of these layers is used for training and validation, and as a result, the average of these results is given. This method gives good results, especially in clustering problems Tarekegn (2020).

Table 6: Cross Validation Classification Report

|  | Precision | Recall | F Score | Support |
|---|---|---|---|---|
| 0 | 0.94 | 0.94 | 0.94 | 1120 |
| 1 | 0.95 | 0.95 | 0.95 | 1335 |
| Accuracy |  |  | 0.94 | 2455 |
| Makro Acc. | 0.94 | 0.94 | 0.94 | 2455 |
| Weighted avg. | 0.94 | 0.94 | 0.94 | 2455 |

It was observed that with layered cross-validation method we achieved 0.94 accuracy, which is same results as the train-test split method and we could use the entire set of 2455 data for training and testing. For this reason, with suitable machine learning methods like KNN, RF, XGB and SVM cross-validation method was carried out; and train-test split method for ANN and LSTM.

## 3.3 Problem and Methodology

Stock market or cryptocurrency price or price movement prediction as a financial time series problem is very important for investors or institutions due to the magnitude of the financial market capitalization. It is very crucial to understand the behavior of the market and solve the mentioned problem to give some recommendations to investors and institutions. Fundamental analysis is the price estimation methodology of financial assets based on economic, social, or political factors. Daily tweets, trend searches, hash-rate, cost calculations, etc., combine fundamental analysis methods to create a better data set. For example, the date the price movements took place, like when a government recognized Bitcoin as currency or when a world-famous company announced that it started using Bitcoin as a trade option, makes a vast Google search number and tweet interaction number that can be taken as a basis for analysis. For this reason, due to the claims of speculation in the Cryptocurrency exchange, daily tweet numbers and Google trend search data were included in the input set and analyzed. The reason for using daily data is to test the models easily with the most commonly used data set and make comparisons with similar previous studies. High-frequency data usage may become possible when obtaining better results from this research.

The main purpose of this study is to measure and compare the prediction performance of the selected ML algorithms. While running the algorithms, continuous and trend data sets are constructed using the technical indicators, tweets, and Google trend data as input features. Also, it tried to test the effect of the number of samples for each data set on prediction performance. That is why reduced data sets were constructed using the last 1-year data for continuous and trend sets. The reason for using complete (2014-2021) and reduced (2020-2021) data sets with mentioned intervals is that they have similar volatility and behavior.

To avoid bias after tabulating calculations of technical indicators, it is essential to perform normalization to ensure the accuracy of model fitting. Recent studies have demonstrated the effectiveness of such data scaling methods in enhancing model performanceAhsan et al. (2021). To mitigate the potential issue of unequal treatment of variables with different scales, MinMax Scalar was employed to normalize the data before model fitting. These normalized data were used in the analysis of continuous data. Considering the data sizes, care was limited to machine learning methods in general, and deep learning methods were not used. Artificial neural networks and recurrent neural networks have been added to machine learning methods and interpreted by comparing training and estimation times, which are not frequently mentioned in other academic studies.

The methodology used in this paper is presented as a model training algorithm 1 as follows.

Seven ML techniques were selected for this study, namely RF, KNN, XGB, SVM, NB, ANN, and LSTM, and they are presented below. During the study, parameter optimization was carried out according to the accuracy test results using trend and continuous data sets for each machine learning technique. Python scikit-learn, Keras, and Tensorflow libraries and their related plugins were used to implement the algorithms. Numpy library was used for mathematical calculations, and Matplotlib and Seaborn libraries were used for visualization and graphics. Previous research conducted in this area was evaluated and reviewed to determine the parameter levels that will be used for the most effective practices. To obtain the best model, the best parameter combination was found for each model separately. All machine learning models were tested using different parameters for each data set according to the best result of the train-test set, we used the same parameters to obtain validation set results. The levels of the parameters for each method are presented in the following subsections.

**Algorithm 1** Model Training Algorithm
---
1. Load the historical data det
2. Create indicators with Ta-Lib
3. Get Tweets and Google Trend Data with Beautiful Soup
4. Binarize data sets to create trend data sets
5. Test ML models for all datasets for best scores
---

### 3.3.1 Random Forest

RF algorithm is a bagging-based machine learning method Breiman (2001). The RF algorithm creates multiple decision trees and combines them to create high-accuracy and stable models. Enlarging the tree in the algorithm increases the randomness rate of the model. RF model, while separating the nodes; instead of the most important feature, looks for the best feature in a random subset of features. This creates diversity and allows better results to be obtained from the model. Ibrahim et al. (2021)

$$Gini = 1 - \sum_{i=1}^{C}(P_i)^2 \qquad (2)$$

$$Entropy = \sum_{i=1}^{C} -P_i * log_2(P_i) \qquad (3)$$

$$Logloss = -\frac{1}{N}\sum_{i=1}^{N}[y_i ln p_i + (1-y_i)ln(1-p_i)] \qquad (4)$$

Table 7: RF parameter levels.

| Parameters | Levels |
|---|---|
| criterion | gini, entropy, log_loss |
| Estimators | 10-500 |
| Max Depth | None,5,10,15 |
| Cross Validation | 20 |

### 3.3.2 K-Nearest Neighbors

The KNN classification method is obtained by measuring the distance in k of the newly entered data to the previously entered data for classification. The formulas used to measure these distances are generally Euclidean5 distance, Minkowski7 distance, and Manhattan7 distance. The most important point of the method is the correct determination of the k value Demir et al. (2019).

$$Euclidean = \sqrt{\sum_{i=1}^{k}(x_i - y_i)^2} \qquad (5)$$

$$Manhattan = \sum_{i=1}^{k}|x_i - y_i| \qquad (6)$$

$$Minkowski = \left(\sum_{i=1}^{k}(|x_i - y_i|)^q\right)^{\frac{1}{q}} \qquad (7)$$

Table 8: KNN parameter levels.

| Parameters | Levels |
|---|---|
| criterion | gini, entropy, log_loss |
| K Range | 1-15 |
| neighbors | None-5 |
| algorithm | balltree, kdtree, brute |
| leaf size | 30 |
| Cross Validation | 20 |

### 3.3.3 Extreme Gradient Boosting

Extreme Gradient Boosting (XGBoost) is an effective and optimized gradient tree boosting algorithmSagi and Rokach (2021). Its high predictive power, avoidance of overfitting, management of missing data, and ability to do all this quickly have made XGBoost one of the most important ML techniques in recent years. It is one of the most efficient algorithms that use decision trees Chen and Guestrin (2016). XGBoost uses a weak learner to minimize the loss function and optimize the objective function (Luo et al., 2021). For any differentiable convex loss function $\imath = \mathbb{R} \times \mathbb{R} \to \mathbb{R}$, the objective function $\mathcal{O}(\phi)$ is defined as:

$$\mathcal{O}(\phi) = \sum_{i=1}^{n} l\left(\hat{y}_i, y_i\right) + \sum_{k} \Omega\left(f_k\right),$$

where $\Omega(f_k) = \gamma T + \frac{1}{2}\lambda \|\omega\|^2$ is a regularization term, $y_i$ are the labels, $w$ is the weight of leaf, and $T$ is the number of leaves of trees.

Table 9: XGB parameter levels.

| Parameters | Levels |
| --- | --- |
| Gradient | 10-500 |
| Max Depth | 3,6,8,10 |
| eval metric | logloss, rmse |
| Cross Validation | 20 |

### 3.3.4 Support Vector Machine

SVM, proposed by Cortes and Vapnik (1995), is a machine learning method based on types of vectors that find a decision region between the two classes that are furthest from any place in the training records. SVMs are mostly used to separate data consisting of two classes (binary classification) but can solve both Classification and Regression problems. Cervantes et al. (2020)

$$\sum_{i \epsilon SV} (a_i - a_i^*) K(x_i, x) + b \qquad (8)$$

Table 10: SVM parameter levels.

| Parameters | Levels |
| --- | --- |
| Kernel | Poly, Linear |
| Degree | 1-5 |
| C range | 0.025,0.1,0.5,1,10,50,100 |
| Gamma | 0.1, 0.5, 1,2,3,4,5 |
| Cross Validation | 20 |

### 3.3.5 Naïve Bayes

NB is a generative learning algorithm widely applied to text classification and sentiment analysis machine learning problems. The classifier is based on the Bayes theorem. This approach to text classification uses the initial feature vector format, where the appearance of a word is modeled by a Gaussian9, Bernoulli11, multinomial10 or Complement12 distribution. Colianni et al. (2015).

$$Gaussian = P(x_i|y) = \frac{1}{\sqrt{2\pi\sigma_y^2}} exp\left(-\frac{(x_i - \mu_y)^2}{2\sigma_y^2}\right) \qquad (9)$$

$$Multinomial = \hat{\theta}_{yi} = \frac{N_{yi} + a}{N_y + an} \qquad (10)$$

$$Bernoulli = P(x_i|y) = P(x_i = 1|y)x_i + (1 - P(x_i = 1|y))(1 - x_i) \qquad (11)$$

$$Complement = w_{ci} = \frac{w_{ci}}{\sum_j |w_{cj}|} \qquad (12)$$

Table 11: NB parameter levels.

| Parameters | Levels |
|---|---|
| Models | Gaussian, Bernoulli, Complement, Multinomial |
| Cross Validation | 20 |

### 3.3.6 Artificial Neural Network

ANN is an algorithm inspired by the human brain's information processing system Cahyani et al. (2021). Learning occurs by processing input and output data and determining the connection weights (weights of the synapses). This process is repeated throughout the system until convergence occurs. Over the past decade, there have been many studies on stock market prediction using artificial intelligence (AI) techniques. These studies initially focused on predicting the return level of the stock price index. In the initial studies, no statistical tests were performed to determine the significance of the empirical results.

Tsaih et al. (1998) integrated ANN with a rule-based technique to predict the direction of change of S&P 500 stock index futures daily. However, stock market data contains tremendous noise and non-stationary; therefore, the ANN training process tends to be difficult Kim (2006).

Table 12: ANN parameter levels.

| Parameters | Levels |
|---|---|
| Activation function | ReLU, tanh, sigmoid |
| Optimizer | SGD, ADAM, NADAM, ADAD, ADAG, FTRL, RMS |
| Loss | SCC, BC |
| Number of hidden layers | 1, 2 |
| Number of hidden neurons | 10, 20, 30, 40, 50 |
| Number of training epochs | 20, 50, 100, 200, 500 |
| Dropout rate | 0.2, 0.3 |

### 3.3.7 Long Short-Term Memory

LSTM is a ML technique based on RNN architecture and learns by remembering inputs at random intervals. When the learning process occurs, the values are stored and do not change. It allows back-and-forth connections between neurons. They are mostly used to process, classify, and predict time series Hochreiter and Schmidhuber (1997). LSTM can be formulated as follows:

$$\mathbf{i}_t = \sigma(\mathbf{W}_{xi}x_t + \mathbf{W}_{hi}\mathbf{h}_{t-1} + \mathbf{W}_{ci}\mathbf{c}_{t-1} + \mathbf{b}_i) \tag{3}$$
$$\mathbf{f}_t = \sigma(\mathbf{W}_{xf}x_t + \mathbf{W}_{hf}\mathbf{h}_{t-1} + \mathbf{W}_{cf}\mathbf{c}_{t-1} + \mathbf{b}_f) \tag{4}$$
$$\mathbf{c}_t = \mathbf{f}_t\mathbf{c}_{t-1} + \mathbf{i}_t \tanh(\mathbf{W}_{xc}x_t + \mathbf{W}_{hc}\mathbf{h}_{t-1} + \mathbf{b}_c) \tag{5}$$
$$\mathbf{o}_t = \sigma(\mathbf{W}_{xo}x_t + \mathbf{W}_{ho}\mathbf{h}_{t-1} + \mathbf{W}_{co}\mathbf{c}_t + \mathbf{b}_o) \tag{6}$$
$$\mathbf{h}_t = \mathbf{o}_t \tanh(\mathbf{c}_t) \tag{7}$$

where $x_t$ and $h_t$ are the input and respectively the hidden state at time $t$, $W_j, W_i$ and $W_0$ are the weight matrices of the corresponding components, $b_f, b_i, b_0$ are the corresponding bias parameters, and $\sigma$ is the activation function.

Table 13: LSTM parameter levels.

| Parameters -Levels | |
|---|---|
| Activation function | ReLU, tanh, sigmoid |
| Optimizer | SGD, ADAM, NADAM, ADAD, ADAG, FTRL, RMS |
| Number of hidden layers | 1, 2 |
| Number of hidden neurons | 10, 20, 30, 40, 50 |
| Number of training epochs | 20,50,100,200,500 |
| Dropout rate | 0.2,0.3 |

## 4 Evaluation criteria

There are some common evaluation metrics to evaluate the models that use only two classes according to the Fawcett (2006). Binary cross entropy loss function was used in Eq 15. As presented in this study, accuracy, F-stat16, and AUC metrics were calculated based on a confusion matrix in Figure 2.

Accuracy is the most commonly used metric for binary classification. The accuracy metric is defined in Eq. 13, and also recall in Eq. 14 based on the elements of the confusion matrix TP, FP, TN, FN illustrated in Figure 2.

$$Accuracy = \frac{(TP + TN)}{(TP + TN + FP + FN)} \tag{13}$$

$$Recall = \frac{(TP)}{(TP + FN)} \tag{14}$$

$$CE = -\frac{1}{n}\sum_{i=1}^{n}\left(y_i \log(p_i) + (1 - y_i)\log(1 - p_i)\right) \qquad (15)$$

$$\sigma^2 = E\frac{TSS - RSS}{p} \qquad (16)$$

The ROC curve is the curve of the recall vs FP rate for all possible threshold values. The AUC is a scalar between 0 and 1. Binary classifier performance increases as the AUC approaches 1.

Figure 2: The components of a standard confusion matrix.

|  |  | Prediction Outcome | | |
|---|---|---|---|---|
|  |  | **p** | **n** | **Total** |
| actual value | **p′** | True Positive | False Negative | P′ |
|  | **n′** | False Positive | True Negative | N′ |
|  | **Total** | P | N |  |

## 5 Results

The experimental result and model comparisons are presented in this section with statistical significance using the paired t-tests at the $p < 0.05$ level. The separation of the continuous and trend data in terms of accuracy was presented in the Appendix 3, 4, 5, 6.

The best performance in all trend data was obtained with the LSTM machine learning method using Nadam optimizer, sparse-categorical-cross-entropy loss function, and 20 epochs with 0.9593 accuracy. The SVM provided 0.9572 accuracy and XGB and Random Forest with 0.9548 accuracy. Naive Bayes method provided the lowest performance with 0.9283.

In trend-reduced data, the best performance was obtained with 0.9421 accuracy of SVM at Regularization = 0.5, Polynomial Degree 3, cross-validation 20 layers setting. XGB was in the second place with an accuracy of 0.9366, while the Random Forest method was in the third place with an accuracy of 0.9339. Artificial Neural Networks showed the lowest accuracy performance with 0.8901 accuracy. Random Forest, XGB, and LSTM reduced trend Dataset training gave the same result as the all trend data set and obtained 0.967 accurate prediction. While the KNN and SVM yielded 0.935, Naive Bayes and Artificial Neural Networks gave 0.903 results. The test set estimation results showed great parallelism with the performance analysis.

With the trend Reduced Data Set, it is observed that the XGB machine learning method, which is very fast in terms of 0.936 accuracy, 0.967 prediction accuracy, and time in train&test data set scores, performs more efficiently in estimating data sets of this size.

When the table for Continuous All Data is examined, the best performance of the Artificial Neural Networks (ANN) machine learning method was obtained with 0.8046 accuracy at the Optimizer = RMSprop, loss: sparse-categorical-crossentropy, epoch = 50 setting. In the second place, the SVM with 0.7943 accuracy, and in the third place with 0.7735, XGB gave the best accuracy. Naive Bayes machine learning method showed the lowest performance with an accuracy rate of 0.6110.

Best Validation Data Set result get with ANN accuracy of 0.8710 among the machine learning methods trained with the Continuous All Data. SVM, Random Forest and XGB Machine learning methods took the second place with an accuracy prediction rate of 0.8388. Naive Bayes method showed the lowest performance with 0.5484 accuracy estimation. With these results, the best and worst estimation rates show parallelism with the performance results. When the training and estimation results are examined with the whole data set, it is thought that the SVM, which is very fast in terms of 0.794 accuracy, 0.8387 estimation accuracy and times for train&test data set, performs more efficiently in estimating data sets of this size.

## 5.1 Continuous datasets

Seven different ML techniques were trained with complete and reduced data sets separately6 5. Estimation models were evaluated on test data sets, and then, to validate the models' performance, the 30-day and 60-day validation data sets were used. The model performance was evaluated using the accuracy, F-stat, and AUC metrics. Paired t-tests were used to compare the best result for each row (shown in bold with a star) with the other results. Results that were not significantly worse ($p > 0.05$) than the best were also shown in bold, representing the top-performing group.

As seen in Table 14, the ANN and SVM models were in the top-performing group with high accuracy, F-stat, and AUC values. The ANN showed the best performance on the complete continuous test data set. These two ML models were significantly better than other estimated ones. This means that the ANN and SVM can be used for BTC price movement for any other data set, such as validation. When looking at Table 15 and 16, the ANN was the best model in 30-day validation data, and the LSTM model was the best in 60-day validation data. The LSTM, RF, and SVM performed comparably to the ANN for 30-day validation data. The SVM, ANN, and LSTM models were in the top-performing group for 30-day validation data sets regardless of which data set was trained on.

For the 60-day estimations, the complete data set trained LSTM and reduced data set trained ANN models showed the best performance with the highest accuracy.

Table 14: Evaluation metrics (test data)

| Models trained using complete data set | RF | KNN | XGBoost | SVM | NB | ANN | LSTM |
|---|---|---|---|---|---|---|---|
| Accuracy | 0.665 | 0.738 | 0.773 | **0.794** | 0.611 | **0.804*** | 0.742 |
| F-stat | 0.678 | 0.772 | 0.792 | **0.817** | 0.655 | **0.832*** | 0.790 |
| AUC | 0.667 | 0.730 | 0.771 | **0.789** | 0.604 | **0.795*** | 0.728 |
| **Models trained using reduced data set** | | | | | | | |
| Accuracy | 0.575 | 0.608 | 0.674 | **0.710** | 0.586 | **0.780*** | 0.681 |
| F-stat | 0.605 | 0.646 | 0.697 | **0.744** | 0.619 | **0.800*** | 0.632 |
| AUC | 0.573 | 0.604 | 0.673 | **0.705** | 0.583 | **0.777*** | 0.676 |

The best accuracy levels obtained from the complete and reduced data-trained ANN models are 0.871 and 0.838 for the 30-day validation set. For the 60-day validation set, these two models are the LSTM and ANN, with a 0.883 accuracy level. Three different evaluation metrics were used to compare the model performance. The best model selected was generally the same for these three metrics. For example, the ANN model outperformed other models for the test data, as seen in Table 14 in terms of accuracy, F-stat, and AUC values.

Table 15: Evaluation metrics (30-day validation data).

| Models trained using complete data set | RF | KNN | XGBoost | SVM | NB | ANN | LSTM |
|---|---|---|---|---|---|---|---|
| Accuracy | **0.838** | 0.733 | 0.774 | **0.838** | 0.548 | **0.871*** | 0.838 |
| F-stat | 0.838 | 0.733 | 0.774 | **0.848** | 0.708 | **0.882*** | 0.857 |
| AUC | **0.846** | 0.736 | 0.781 | **0.840** | 0.500 | **0.869*** | 0.843 |
| **Models trained using reduced data set** | | | | | | | |
| Accuracy | 0.645 | 0.613 | 0.741 | **0.806** | 0.646 | **0.838*** | 0.800 |
| F-stat | 0.685 | 0.625 | 0.733 | **0.812** | 0.620 | **0.857*** | 0.833 |
| AUC | 0.638 | 0.615 | 0.752 | **0.810** | 0.657 | **0.834*** | 0.799 |

It is possible to make some general comments by examining the results of the analysis. The sample size significantly affects the prediction performance by decreasing the accuracy level. When the estimated models are examined, it can be easily seen that the accuracy level decreases in reduced models. One of the most important results obtained from this study is the ANN model showed significantly better results for small samples. The results, especially the 30-day and 60-day validation data prediction, showed that the estimated ML models did not become overfit, and these models can be used for the selected problem properly.

Table 16: Evaluation metrics (60-day validation data).

| Models trained using complete data set | RF | KNN | XGBoost | SVM | NB | ANN | LSTM |
|---|---|---|---|---|---|---|---|
| Accuracy | 0.833 | 0.667 | 0.833 | **0.867** | 0.567 | 0.833 | **0.883*** |
| F-stat | 0.848 | 0.667 | 0.848 | **0.882** | 0.717 | 0.844 | **0.898*** |
| AUC | 0.840 | 0.686 | 0.840 | **0.869** | 0.491 | 0.846 | **0.882*** |
| **Models trained using reduced data set** | | | | | | | |
| Accuracy | 0.800 | 0.700 | 0.766 | **0.867** | 0.616 | **0.883*** | 0.750 |
| F-stat | 0.823 | 0.743 | 0.788 | **0.882** | 0.657 | **0.904*** | 0.819 |
| AUC | 0.800 | 0.691 | 0.771 | **0.869** | 0.614 | **0.871*** | 0.706 |

## 5.2 Trend datasets

Seven different ML techniques 5 6 3 4 were trained using complete and reduced data sets same as using the continuous data set procedures. In this section, trend data sets were constructed and estimated models were evaluated and compared. Paired t-tests were used to compare the best result for each row (shown in bold with a star) with the other results. Results that were not significantly worse (p > 0.05) than the best were also shown in bold, representing the top-performing group.

As can be seen in Tables 17, 18, and 19 using trend data set improved the prediction performance of the selected techniques. When comparing the continuous data set results with the trend data set, it can be seen easily the higher level of accuracy, F-stat, and AUC values. Also, the accuracy values or F-stat and AUC are close to each other for each estimated ML model, which is trained using trend data sets. This means that converting the continuous data to trend data sets made the problem easier to predict and the ML techniques showed similar performance with little differences. The LSTM model trained using the complete data set showed the best performance on test data with the highest accuracy level of 0.959, with the ANN, SVM, XGBoost, and RF being in the top-performing group that were not significantly worse than the best model for the test data set. Usage of the reduced trend data decreased the accuracy level, and interestingly, the ANN and LSTM models were not in the top-performing group. The SVM showed the best performance with the RF and XGBoost models with no significant differences in terms of accuracy, F-stat, and AUC.

Table 17: Evaluation metrics (test data)

| Models trained using complete data set | RF | KNN | XGBoost | SVM | NB | ANN | LSTM |
|---|---|---|---|---|---|---|---|
| Accuracy | **0.954** | 0.944 | **0.954** | **0.957** | 0.928 | **0.951** | **0.959*** |
| F-stat | **0.958** | 0.948 | **0.959** | **0.960** | 0.935 | **0.954** | **0.962*** |
| AUC | **0.954** | 0.945 | **0.953** | **0.957** | 0.927 | **0.954** | **0.960*** |
| **Models trained using reduced data set** | | | | | | | |
| Accuracy | **0.933** | 0.911 | **0.937** | **0.942*** | 0.900 | 0.890 | 0.923 |
| F-stat | **0.938** | 0.917 | **0.940** | **0.945*** | 0.908 | 0.903 | 0.925 |
| AUC | **0.934** | 0.912 | **0.938** | **0.944*** | 0.901 | 0.884 | 0.922 |

For 30-day prediction, all estimated ML models provided no significant differences with almost equal accuracy, F-stat, and AUC for the models trained using the complete data sets. The LSTM model showed the best performance with a 0.967 level of accuracy. For 60-day data prediction models gave completely different results. The KNN outperformed all other estimated models with a 0.967 level of accuracy and also with statistically significant differences.

Table 18: Evaluation metrics (30 days validation data)

| Models trained using complete data set | RF | KNN | XGBoost | SVM | NB | ANN | LSTM |
|---|---|---|---|---|---|---|---|
| Accuracy | **0.965** | **0.965** | 0.960 | **0.966** | 0.959 | **0.961** | **0.967*** |
| F-stat | **0.965** | **0.966** | 0.964 | **0.967** | 0.960 | 0.959 | **0.969*** |
| AUC | **0.965** | **0.966** | 0.964 | **0.966** | 0.960 | 0.959 | **0.970*** |
| **Models trained using reduced data set** | | | | | | | |
| Accuracy | **0.967*** | 0.935 | **0.967*** | 0.935 | 0.903 | 0.903 | **0.967*** |
| F-stat | **0.969*** | 0.937 | **0.969*** | 0.937 | 0.903 | 0.909 | **0.969*** |
| AUC | **0.970*** | 0.941 | **0.970*** | 0.941 | 0.912 | 0.905 | **0.970*** |

The performance of the models trained with reduced data sets was generally get lower results than the complete data sets in terms of selected evaluation metrics. The SVM model outperformed other estimated ML models for test data with a 0.942 accuracy, however, the RF and XGBoost models were in the top-performing group with no significant differences. The LSTM, XGBoost, and RF models showed the best prediction performance with a 0.967 level of accuracy and outperformed the other 4 ML models for the 30-day validation data. The same models were in the top-performing group but had lower accuracy, as expected for predicting 60-day validation data.

Table 19: Evaluation metrics (60 days validation data)

| Models trained using complete data set | RF | KNN | XGBoost | SVM | NB | ANN | LSTM |
|---|---|---|---|---|---|---|---|
| Accuracy | 0.950 | **0.967*** | 0.950 | 0.950 | 0.933 | 0.950 | 0.950 |
| F-stat | 0.957 | **0.978*** | 0.957 | 0.957 | 0.938 | 0.955 | 0.957 |
| AUC | 0.951 | **0.978*** | 0.951 | 0.951 | 0.943 | 0.957 | 0.951 |
| **Models trained using reduced data set** | | | | | | | |
| Accuracy | **0.950*** | 0.917 | **0.950*** | 0.900 | 0.883 | 0.933 | **0.951*** |
| F-stat | **0.957*** | 0.923 | **0.957*** | 0.909 | 0.888 | 0.939 | **0.959*** |
| AUC | **0.951*** | 0.929 | **0.951*** | 0.908 | 0.900 | 0.942 | **0.959*** |

## 5.3 Model comparison for training and testing

Model accuracy is crucial, but it is not the only metric for choosing the best one to make more profit. More time-consuming models would not be suitable for financial investors many times because sometimes even seconds gain great importance in stock market or cryptocurrency trading transactions. For this reason, the prediction method must be chosen correctly to make the right decision, in other words, to make more money or profit.

As seen in table 20 and 21, calculated elapsed time values may look like they are not important due to their magnitude, but if a system uses a big data set containing thousands of features and billions of samples may need a lot of time to provide accurate predictions. The most important thing at this point is to find more suitable models regarding accuracy and elapsed time.

Table 20: Elapsed time (Continuous Data Set)

| | Complete Data Set | | Reduced Data Set | |
|---|---|---|---|---|
| | Training Time | Prediction Time | Training Time | Prediction Time |
| LSTM | 33,972134 | 3,253587 | 49,599901 | 2,938089 |
| ANN | 7,369106 | 0,169317 | 14,201830 | 0,464278 |
| RF | 4,980866 | 0,072236 | 0,056011 | 0,008095 |
| SVM | 0,427818 | 0,007975 | 0,024007 | 0,008005 |
| XGBoost | 0,120009 | 0,008000 | 0,032001 | 0,008003 |
| KNN | 0,008016 | 0,015995 | 0,008002 | 0,016004 |
| NB | 0,008000 | 0,008002 | 0,008004 | 0,005010 |

Regarding train&test and validation results for the complete continuous data set, the SVM machine learning method, with 0.794 accuracy for train&test and 0.8387 for validation accuracy, performed more efficiently in predicting data sets. When the training and prediction times for the complete trend data sets are examined in terms of accuracy; It is thought that the SVM method performs more efficiently than other methods. For the training and prediction results with the reduced trend data set, it was concluded that the XGBoost machine learning method, which is very fast in terms of time, provided 0.936 accuracy for train&test and 0.967 for validation accuracy.

For the reduced continuous data set, with 0.78 accuracy for train&test and 0.839 validation accuracy, ANN performance stands out among other ML methods. However, in terms of time, it is 500 times slower than SVM, which has the closest performance of 0.71 accuracy for train&test and 0.80 validation accuracy. However, ANN is the best option for processing reduced data in a continuous structure. Among the ML models, all models except the LSTM model gave fast results for prediction in less than 1 second, while in terms of the time required for training, models other than LSTM and ANN generally achieved speeds below 1 second.

The LSTM and ANN models need more time for training for both continuous and trend data sets. These models need more data points for training because of the mathematical structure and complexity. A small sample size is not suitable for these models. The LSTM and ANN techniques are mentioned as deep learning methods that use big data in many recent studies, such as Nita (2016), Wang et al. (2020), Sismanoglu et al. (2019).

Table 21: Elapsed time (Trend Data Set)

|  | Complete Data Set | | Reduced Data Set | |
| --- | --- | --- | --- | --- |
|  | Training Time | Prediction Time | Training Time | Prediction Time |
| LSTM | 32,262309 | 0,120025 | 32,331675 | 2,616832 |
| ANN | 7,879686 | 0,194058 | 3,611537 | 0,162802 |
| RF | 0,415293 | 0,032011 | 0,413021 | 0,030233 |
| SVM | 0,360201 | 0,008007 | 0,008004 | 0,008002 |
| XGBoost | 0,402129 | 0,016006 | 0,032004 | 0,008001 |
| KNN | 0,032311 | 0,008003 | 0,008003 | 0,008005 |
| NB | 0,008005 | 0,007996 | 0,800000 | 0,008001 |

# 6 Conclusions and Discussion

In the world of trading, prediction price direction has historically been important. Thanks to its trading volume, Bitcoin, the first of the cryptocurrencies, is frequently analyzed by researchers and investors. In this study, in addition to the indicators frequently used in the financial market for Bitcoin price prediction, daily Google Search Trends and the number of daily tweets on Bitcoin are also included in the input feature space. Seven ML techniques were tested for the prediction of price movement using continuous and trend data sets. To obtain the best parameter set for each ML technique, specific parameter levels were used as presented in section 3.3. The optimized settings for trend and continuous data were almost different for each data set. Therefore, the importance of optimizing the method according to the structure of the data instead of a general setting has once again emerged in the analysis.

It was observed in the significance charts that the impact of the number of daily tweets about Bitcoin and the change in Google Search Trends compared to the previous day is not as decisive as the indicators. For this reason, new data generated with indicators, such as moving mediums for these data, can be added to other technical indicators.

In the performance tests of selected ML techniques using continuous data, there were significant differences between the accuracy of the models. While the highest accuracy of 0.804 was achieved with the ANN model for train&test scores among models for complete continuous data, accuracy dropped to 0.611 in models such as Naive Bayes. For reduced continuous data, the best accuracy level reached 0.780 with the ANN, and the lowest was 0.575 with the RF. In continuous data sets, prediction performances were similar for both train&test and validation data set. It has been observed that the estimated rates of reduced test data for the RF model are reduced by approximately 16%. The ANN model showed the best prediction performance for all data sets except for 60-day validation data. For 60-day validation data, LSTM showed the best performance. However, the calculation time and processing power required for neural networks for training and prediction are considerably higher than the models other than LSTM. SVM and XGBoost models provided efficient results regarding running time and performance. The results obtained from trend data showed that the rapid changes in price movements in the reduced data provide enough data to build a model.

Neural network models like ANN and LSTM are known to perform well in price movement direction and price prediction. For this reason, it is thought that the models created instead of static single predictions may provide better performance using Pipeline models. Dong et al. (2020)

As the transaction volume of cryptocurrency markets increases, price and price movement prediction models will continue to evolve.In this research, various ML models were tested in terms of trend and continuous data to predict BTC price movement, and it was understood that good results can be obtained with less data. These prediction models can be used for other financial time series data models with some improvements and revisions. Since cryptocurrency exchanges stream continuously on a 7-24 basis, instant trading is more important than daily closing prices. For this reason, we believe that the models from which we get efficient results will also effectively estimate different frequency data sets such as hours, minutes, and 15 seconds.

The information obtained from the estimated models can help investors take short/long-term positions, especially in future markets, and develop novel models for buy-hold-sell signal generation. Since deep learning models give better results with larger data, they may provide better predictions and bring the studies to a higher level. The text classification models may also provide better results by combining the sentiments with financial data. With the approval of spot Bitcoin Exchange Traded Funds by the US Securities and Exchange Commission on January 10, 2024, it is thought that studies in this field will increase rapidly.

**Funding** We declare that we have no relevant or material financial interests that relate to the research described in this paper. The authors declare that no funds, grants, or other support were received during the preparation of this manuscript.

# 7  Declarations

Conflict of interest The authors have not disclosed any conflict of interest.

Madan, I., Saluja, S., and Zhao, A. (2015). Automated bitcoin trading via machine learning algorithms. *URL: http://cs229. stanford. edu/proj2014/Isaac% 20Madan*, 20.

Mangla, N., Bhat, A., Avabratha, G., and Bhat, N. (2019). Bitcoin price prediction using machine learning. *International Joutnal of Information And Computing Science*, 6(5):318–320.

Nita, S. (2016). Machine learning techniques used in big data. *Scientific Bulletin of Naval Academy*, 19.

Pabuçcu, H., Ongan, S., and Ongan, A. (2020). Forecasting the movements of bitcoin prices: an application of machine learning algorithms. *Quantitative Finance and Economics*, 4(4):679–692.

Rahaman, A., Bitto, A. K., Biplob, K., Bijoy, M. H. I., Jahan, N., and Mahmud, I. (2022). Bitcoin trading indicator: a machine learning driven real time bitcoin trading indicator for the crypto market. *Bulletin of Electrical Engineering and Informatics*, 12:1762–1772.

Sagi, O. and Rokach, L. (2021). Approximating xgboost with an interpretable decision tree. *Information Sciences*, 572:522–542.

Shah, D., Isah, H., and Zulkernine, F. (2019). Stock market analysis: A review and taxonomy of prediction techniques. *International Journal of Financial Studies*, 7(2):26.

Shah, D. and Zhang, K. (2014). Bayesian regression and bitcoin. In *2014 52nd annual Allerton conference on communication, control, and computing (Allerton)*, pages 409–414. IEEE.

Sismanoglu, G., Onde, M. A., Kocer, F., and Sahingoz, O. K. (2019). Deep learning based forecasting in stock market with big data analytics. pages 1–4.

Sossi-Rojas, S., Velarde, G., and Zieba, D. (2023). A machine learning approach for bitcoin forecasting. *Engineering Proceedings*, 39(1).

Tarekegn, A. N. (2020). Cross-validation approach to evaluate clustering algorithms: An experimental study using multi-label datasets. *SN Computer Science*, 1(5):263.

Tsaih, R., Hsu, Y., and Lai, C. C. (1998). Forecasting sp 500 stock index futures with a hybrid ai system. *Decision Support Systems*, 23(2):161–174.

Wang, C., Du, W., Zhu, Z., and Yue, Z. (2020). The real-time big data processing method based on lstm or gru for the smart job shop production process. *Journal of Algorithms & Computational Technology*, 14:1748302620962390.
29

# 8 Appendix

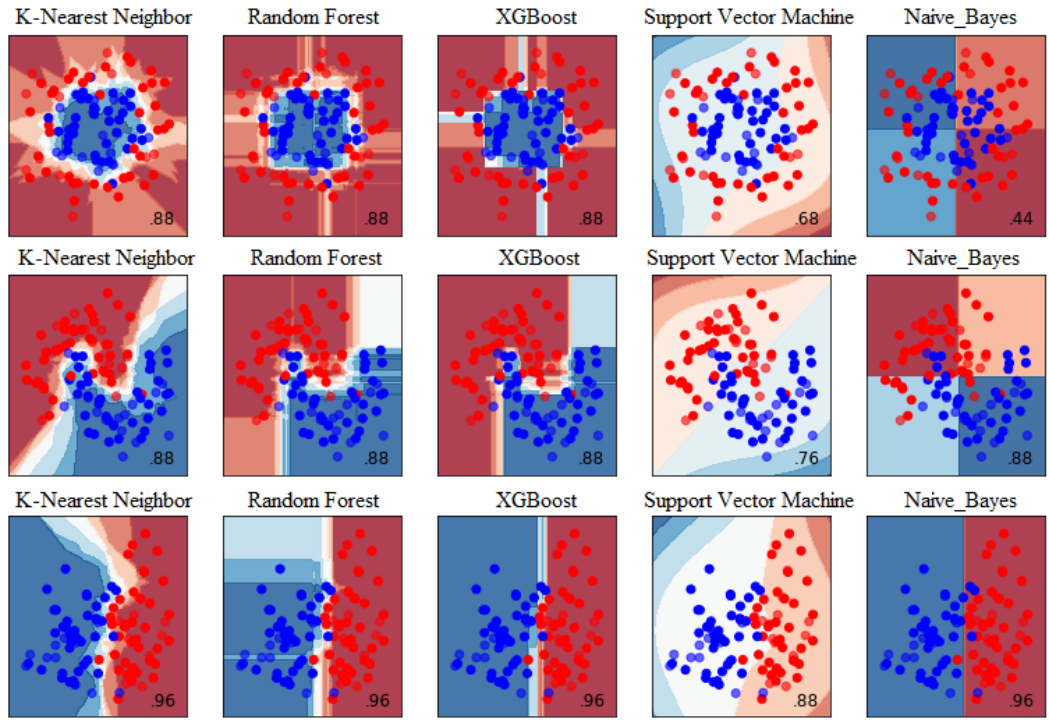

Figure 3: Separation of all trend data according to accuracy

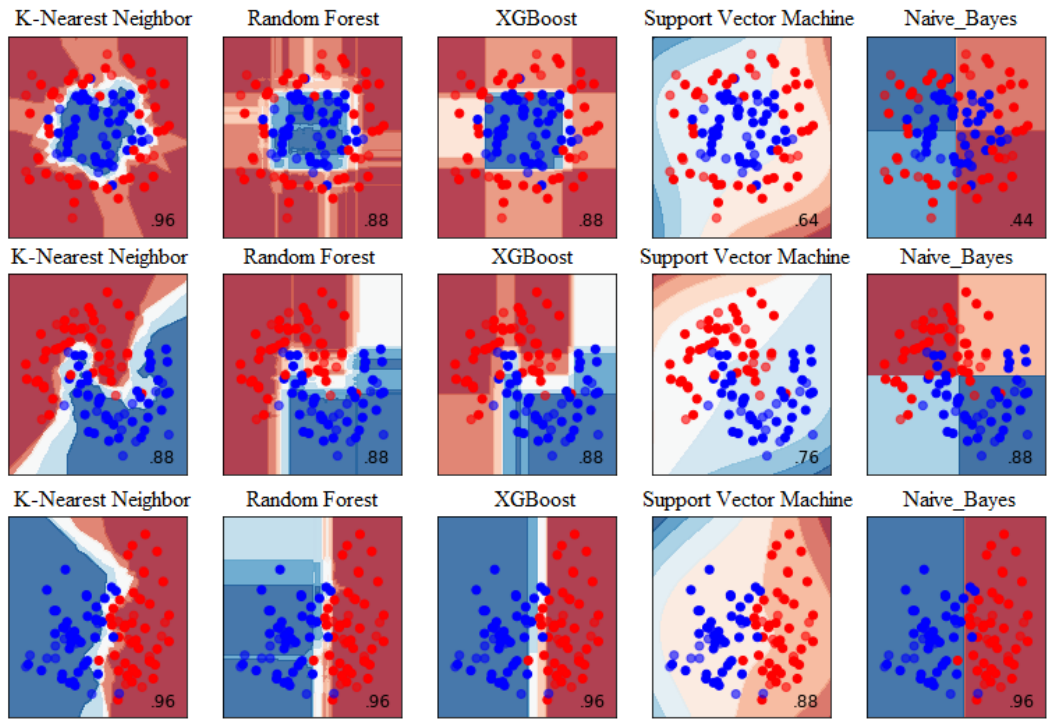

Figure 4: Separation of 1 year trend data according to accuracy

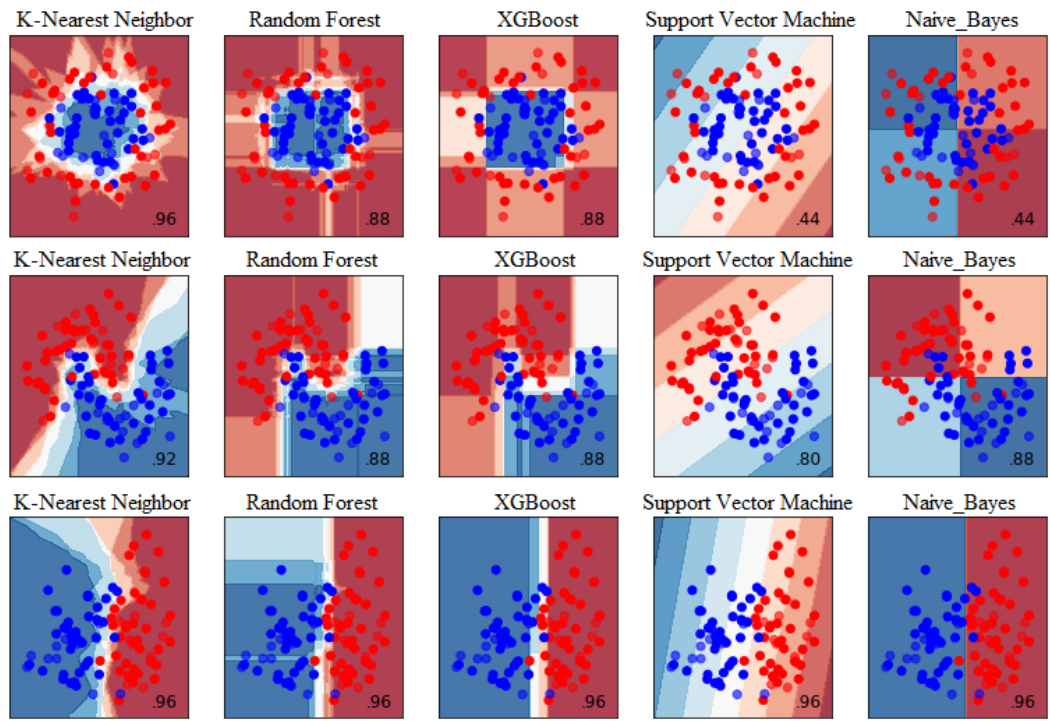

Figure 5: Separation of all continuous data according to accuracy

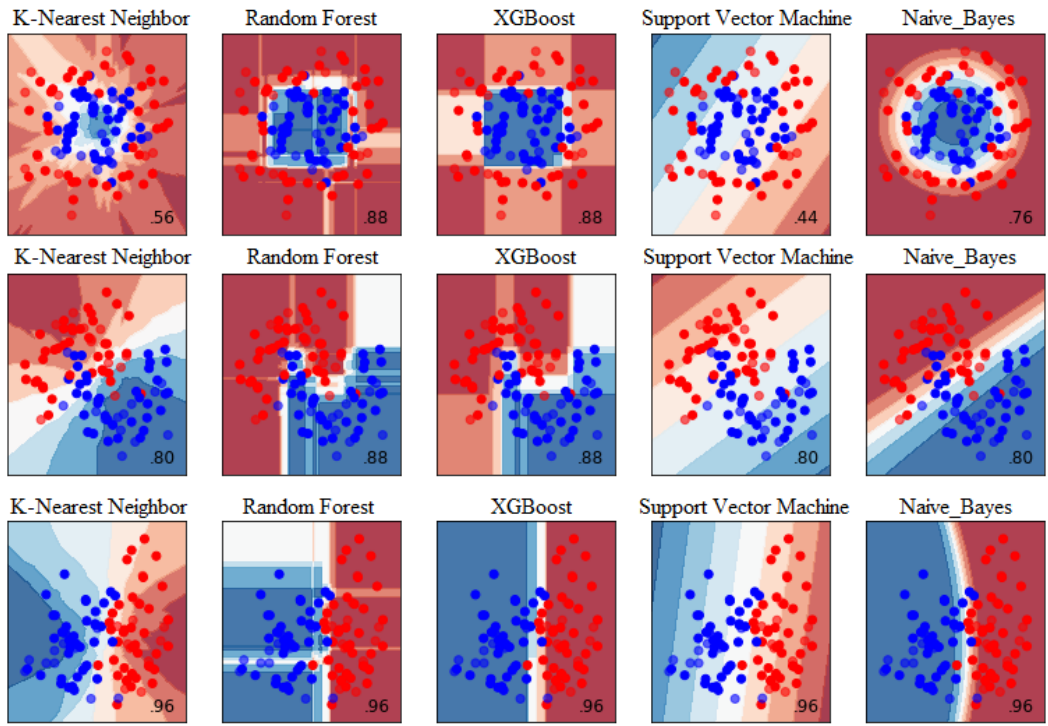

Figure 6: Separation of 1 year continuous data according to accuracy